\documentclass[12pt]{article}
\usepackage{amsmath}
\def\g{\gamma}

\def\be{\begin{equation}}
\def\ee{\end{equation}}
\def\arr{\begin{array}{rll}}
\def\ea{\end{array}}
\def\bea{\begin{eqnarray}}
\def\eea{\end{eqnarray}}

\def\N2{$N{=}2$}

\def\>{\rangle}
\def\<{\langle}
\def\+{\dagger}
\def\={\ =\ }

\begin{document}
\renewcommand{\thefootnote}{\fnsymbol{footnote}}
\begin{titlepage}
\setcounter{page}{0}
\vskip 1cm
\begin{center}
{\LARGE\bf Geometry of the isotropic oscillator }\\
\vskip 0.5cm
{\LARGE\bf driven by the conformal mode }\\
\vskip 1cm
$
\textrm{\Large Anton Galajinsky \ }
$
\vskip 0.7cm
{\it
School of Physics, Tomsk Polytechnic University,
634050 Tomsk, Lenin Ave. 30, Russia} \\
{e-mail: galajin@tpu.ru}

\end{center}
\vskip 1cm
\begin{abstract} \noindent
Geometrization of a Lagrangian conservative system typically amounts to reformulating its equations of motion as the geodesic equations in a properly chosen curved spacetime. The conventional methods include the Jacobi metric and the Eisenhart lift. In this work, a modification of the Eisenhart lift is proposed which describes the isotropic oscillator in arbitrary dimension driven by the one-dimensional conformal mode.
\end{abstract}

\vskip 1cm
\noindent
Keywords: Eisenhart lift, Newton--Hooke symmetry

\end{titlepage}

\renewcommand{\thefootnote}{\arabic{footnote}}
\setcounter{footnote}0

\noindent
{\bf 1. Introduction}\\

\noindent
It would not be an exaggeration to say that, since the discovery of general relativity, geometry and theoretical physics go parallel.
Given a dynamical system with finitely many degrees of freedom, it is customary to ask: What is its geometric description? Vice versa, given a geometric idea, it is natural to wonder: What would be its physical application?

In general, geometrization of a Lagrangian conservative system, whose kinetic term involves a positive definite metric, amounts to reformulating its equations of motion as the geodesic equations in a properly chosen curved spacetime or embedding them into the geodesics of a larger theory such that the time evolution of the extra degrees of freedom is unambiguously fixed, provided the dynamics of the original model is known. The Jacobi metric \cite{Ong} and the Eisenhart lift \cite{Eis,duv} represent the conventional tools of that kind (for a recent application to time--dependent systems see \cite{CDGH}). While the former method yields a Riemannian metric, the latter gives a Lorentzian metric thus paving the way for applications in the general relativistic context.

The goal of this work is to discuss a geometric formulation for a particular dynamical system which describes the isotropic oscillator in arbitrary dimension driven by the one--dimensional conformal mode (see Eq. (\ref{Osc}) below). It naturally arises if one applies the conventional method of non--linear realizations \cite{CWZ} to the conformal extension of the Newton--Hooke algebra \cite{GM}.~\footnote{It is customary to refer to this algebra as the $l=\frac 12$ conformal Newton--Hooke algebra. Note that the equations of motion similar to (\ref{Osc}) were first obtained in \cite{FIL} when constructing dynamical realizations for the so called $l=1$ conformal Galilei algebra. As compared to (\ref{Osc}), the $\frac{1}{L^2}$ term was absent and the frequency of oscillation was bigger by a factor of two. A geometric formulation for the rightmost equation in (\ref{Osc}) (with no $\frac{1}{L^2}$ term) was first proposed in \cite{IKL}. An alternative parametrization of the coset space for the $l$--conformal Newton--Hooke algebra was recently discussed in \cite{KLS}.} The model describes a particle moving along the ellipse such that the periods of its accelerated or decelerated motion are controlled by a single function of time variable representing the $d=1$ conformal mechanics of de Alfaro, Fubini, and Furlan \cite{DFF}. It is demonstrated below that, by properly including the conformal mode into the Eisenhart metric associated with the isotropic oscillator, one can attain a satisfactory geometric description in terms of a Lorentzian metric which solves the Einstein equations.

The organization of the paper is as follows: In Sect. 2, the isotropic oscillator in arbitrary dimension driven by the $d=1$ conformal mode is discussed. Its symmetries and the general solution to the equations of motion are given. Sect. 3 contains a brief account of the conventional Eisenhart lift applied to conservative mechanical systems. An embedding of the isotropic oscillator driven by the one--dimensional conformal mode into the geodesics of the Eisenhart--like metric is discussed in Sect. 4.
The conformal mode enters the metric as a scale factor in such a way that the oscillator equation follows from the geodesics, while the conformal mechanics arises when imposing
the Einstein equations. In Sect. 5, the Killing isometries of the metric are studied. A set of vector fields is found which all together form the Newton--Hooke algebra under the commutator.
The requirement that they be the Killing vector fields of the metric in Sect. 4 reproduces the conformal mechanics equation on the scale factor. It is demonstrated that the same condition occurs if one demands  the spacetime to be stationary. We summarize our results and discuss possible further developments in the concluding Sect. 6. Throughout the paper summation over repeated indices is understood.

\vspace{0.5cm}

\noindent
{\bf 2. Isotropic oscillator driven by $d=1$ conformal mode}\\

\noindent
Consider the configuration space $\mathbf{R}^1 \times \mathbf{R}^d$ parametrized by the coordinates $\rho$ and $x_i$, $i=1,\dots,d$, and the dynamical system governed by the equations of motion
\be\label{Osc}
\rho(t)^2 \frac{d}{dt} \left(\rho(t)^2 \frac{d}{dt} x_i(t) \right)+\gamma^2 x_i(t)=0, \qquad \frac{d^2 \rho(t)}{dt^2}=\frac{\gamma^2}{\rho(t)^3}-\frac{\rho(t)}{L^2},
\ee
where $\gamma$ and $L$ are positive constants. The rightmost equation describes the $d=1$ conformal mechanics \cite{DFF}, while the leftmost equation corresponds to the isotropic oscillator in $d$ dimensions driven by the conformal mode.

It is straightforward to verify that Eqs. (\ref{Osc}) hold invariant under the infinitesimal transformations
\bea\label{tr}
&&
t'=t+a-b L^2 (\cos{(2t/L)}-1)+c L \sin{(2t/L)},
\nonumber\\[4pt]
&&
\rho'(t')=\rho(t)+\left(b L \sin{(2t/L)}+c \cos{(2t/L)}\right) \rho(t),
\nonumber\\[4pt]
&&
x'_i (t')=x_i (t)+\frac{1}{\rho(t)} \cos{(t/L)} \alpha_i+\frac{1}{\rho(t)} L \sin{(t/L)} \beta_i+\omega_{ij} x_j(t),
\eea
which provide a realization of the conformal extension of the Newton--Hooke group \cite{GM}. The parameters $(a,b,c,\alpha_i,\beta_i,\omega_{ij}=-\omega_{ji})$ are associated with the time translation, special conformal transformation, dilatation, spatial translations, Newton--Hooke boosts, and spatial rotations, respectively, while $L$ is identified with the characteristic time \cite{BL}.
In order to verify the structure relations of the conformal Newton--Hooke algebra, it suffices to consider the variations of fields
\be
\delta \rho(t)=\rho'(t)-\rho(t), \qquad \delta x_i(t)=x'_i(t)-x(t),
\ee
and compute their commutators.\footnote{When evaluating the commutators, it is to be remembered that the variations act upon the fields $\rho(t)$ and $x_i(t)$ and
do not affect the temporal coordinate $t$.}

Making use of the conserved charges associated with the symmetry transformations (\ref{tr}), one can then build the general solution to (\ref{Osc}) by purely algebraic means
\be\label{sol}
\qquad \qquad \quad ~
\rho(t)=\sqrt{\frac{\left( \mathcal{D} L\sin{(t/L)}+\mathcal{C} \cos{(t/L)}\right)^2+{\left( \g L \sin{(t/L)} \right)}^2}{\mathcal{C}}},
\ee
\be\label{sol1}
x_i(t)=\frac{1}{\rho(t)} L \sin{(t/L)} \mathcal{P}_i-\frac{1}{\rho(t)} \cos{(t/L)} \mathcal{K}_i,
\ee
where $\mathcal{D}$, $\mathcal{C}$, $\mathcal{P}_i$, $\mathcal{K}_i$ are constants of integration.

The shape of the trajectory in the Newton--Hooke reference frame is readily found if one uses the alternative parametrization of the curve
\be\label{s}
\rho(t)^2 \frac{d}{dt}=\frac{d}{d\varphi}, \qquad  \frac{d\varphi}{dt}=\frac{1}{\rho^2}, \qquad
\varphi(t)=\frac{1}{\g}\arctan{\frac{\mathcal{D}\mathcal{C}+(\mathcal{D}^2+\g^2) L\tan{(t/L)}}{\g\mathcal{C}}},
\ee
which turns the leftmost equation in (\ref{Osc}) into that describing an ordinary isotropic oscillator. The orbit is thus an ellipse
\be\label{CHI}
x_i (t)=\mu_i \cos{(\g \varphi(t))}+\nu_i \sin{(\g\varphi(t))},
\ee
$\mu_i$ and $\nu_i$ being constants of integration.
According to (\ref{sol}), the conformal mode is an oscillating function which determines the periods of accelerated/decelerated motion of the particle along the ellipse. Most easily this is illustrated by examining the circular motion ($\mu^2=1$, $\nu^2=1$, $(\mu,\nu)=0$) for which
\be
|\dot {\vec{x}}|=\frac{\gamma}{\rho(t)^2}.
\ee
The increase/decrease of $\rho(t)$ causes the deceleration/acceleration of a particle along the ellipse.

Worth mentioning is that the minimum point of the potential $U(\rho)=\frac{\gamma^2}{\rho^2} +\frac{\rho^2}{L^2}$, which occurs at
\be\label{ps}
\rho_0=\sqrt{\gamma L},
\ee
provides a particular solution to the rightmost equation in (\ref{Osc}) and turns the leftmost equation into
\be\label{OSS}
\frac{d^2  x_i(t)}{dt^2}+\frac{1}{L^2} x_i(t)=0.
\ee
An ordinary isotropic oscillator is thus a particular instance of (\ref{Osc}). Eq. (\ref{OSS}) also arises if one goes over to the non--inertial Newton--Hooke reference
frame\footnote{This reference frame also indicates the Lagrangian formulation
$\mathcal{L}=\left({\dot\rho}^2-\frac{\gamma^2}{\rho^2}-\frac{\rho^2}{L^2}\right)+\left({(\rho x_i)}^{.}  {(\rho x_i)}^{.} -\frac{\rho^2 x_i x_i}{L^2}\right)$ which reproduces the equations of motion (\ref{Osc}). We thank S. Krivonos for pointing this out to us. Note that the Eisenhart metric associated with such a Lagrangian does not solve the Einstein equations unless $\rho$ takes a fixed value as in (\ref{ps}).}
\be\label{TRA}
x'_i=\rho(t) x_i,
\ee
which correlates with Eq. (\ref{sol1}) above.

\vspace{0.5cm}

\noindent
{\bf 3. The Eisenhart lift}\\

\noindent
The Eisenhart lift \cite{Eis,duv} suggests a geometric formulation for a conservative mechanical system with $d$ degrees of freedom $x_i$, $i=1,\dots,d$, and the potential energy $U(x)$ in terms of geodesics of the Lorentzian metric
\be\label{metric}
g_{\mu\nu}(y) dy^\mu dy^\nu=-2 U(x) d t^2-dt ds+dx_i dx_i
\ee
defined on a $(d+2)$--dimensional spacetime parametrized by $y^\mu=(t,s,x_i)$, where $t$ is identified with the temporal variable of the Newtonian mechanics and $s$ is an extra coordinate.

Computing the Christoffel symbols
\be\label{cs}
\Gamma^i_{tt}=\partial_i U, \qquad \Gamma^s_{t i}=2\partial_i U,
\ee
where we have split the index $\mu=(t,s,i)$, $i=1,\dots,d$, and abbreviated $\partial_\mu=\frac{\partial}{\partial y^\mu}$, and analyzing the geodesic equations
\be
\frac{d^2 y^\lambda}{d \tau^2}+\Gamma^\lambda_{\mu\nu} \frac{d y^\mu}{d \tau}  \frac{d y^\nu}{d \tau}=0, \qquad g_{\mu\nu} \frac{d y^\mu}{d \tau} \frac{d y^\nu}{d \tau}=\epsilon,
\ee
where $\epsilon=0$ for the null geodesics and $\epsilon=-1$ for the time--like geodesics, one concludes that
$t$ is affinely related to the proper time $\tau$
\be\label{lt}
\frac{d^2t}{d \tau^2}=0 \quad \Rightarrow \quad \frac{d t}{d \tau}=\kappa,
\ee
where $\kappa$ is a positive constant, while $x_i$ obeys Newton's equation (passing from $\tau$ to $t$ via (\ref{lt}))
\be\label{EQN}
\frac{d^2 x_i}{d t^2}+\partial_i U=0.
\ee
Given the general solution to (\ref{EQN}), the dynamics of $s$ is fixed from the condition that the geodesic is null or time--like (again passing from $\tau$ to $t$)
\be\label{eqs1}
\frac{d s}{d t}=\frac{d x_i}{dt} \frac{d x_i}{dt}-2 U-\frac{\epsilon}{\kappa^2},
\ee
The conventional Newtonian mechanics is thus recovered by implementing the null reduction along $s$ \cite{Eis,duv}.

A remarkable property of the Eisenhart metric is that it holds invariant under the transformation
\be
s'=s+\lambda,
\ee
where $\lambda$ is a constant,
which gives rise to the covariantly constant null Killing vector field
\be\label{Xi}
\xi^\mu \partial_\mu=\partial_s, \qquad \nabla_\mu \xi_\nu=0, \qquad \xi^2=0.
\ee
The latter determines the so called Bargmann structure on the manifold \cite{duv}.
Eq. (\ref{Xi}) also implies that the spacetime (\ref{metric}) admits a geodesic null congruence with vanishing expansion, shear and vorticity and thus belongs to the Kundt class.

The distinguished vector field $\xi^\mu$ can be used to construct the trace--free energy--momentum tensor\footnote{The factor $\frac{d}{2\pi}$, $d$ being the dimension of the $x$--subspace, is chosen for further convenience.}
\be\label{EMT}
T_{\mu\nu}=\frac{d}{2\pi} \Omega(y)^2 \xi_\mu \xi_\nu, \qquad {T^\mu}_\mu=0,
\ee
where $\Omega(y)^2$ is an arbitrary function (the energy density). Because the only non--vanishing component of $\xi_\mu$ is $\xi_t=-\frac 12$,
the energy--momentum tensor has only one nonzero component
\be
T_{tt}=\frac{d }{8\pi} \Omega^2.
\ee

Taking into account that $\xi_\mu$ is covariantly constant, one can compute the covariant derivative
\be
\nabla_\rho T_{\mu\nu}=\frac{d}{\pi} \Omega \partial_\rho \Omega\xi_\mu \xi_\nu,
\ee
and verify that the energy--momentum tensor is conserved
\be
\nabla^\mu T_{\mu\nu}=0,
\ee
provided $\Omega$ does not depend on $s$, i.e. $\Omega=\Omega(t,x)$.

Finally, making use of the Christoffel symbols (\ref{cs}), one finds that the Ricci tensor has only one nonzero component $R_{tt}$ while the scalar curvature vanishes
\be
R_{tt}=\partial_i \partial_i U, \qquad R=0.
\ee
Given the Eisenhart metric (\ref{metric}) and the energy--momentum tensor (\ref{EMT}), the conventional Einstein equations
\be\label{EE}
R_{\mu\nu}-\frac 12 g_{\mu\nu}(R+2\Lambda)=8\pi T_{\mu\nu}
\ee
imply that the contribution of the cosmological term necessarily vanishes $\Lambda=0$ thus reducing (\ref{EE}) to
\be\label{EE1}
R_{\mu\nu}=8\pi T_{\mu\nu},
\ee
only $(tt)$--component of which is non--trivial.

A particularly interesting example of the Eisenhart geometry occurs it one takes $\Omega^2$ in (\ref{EMT}) to be $s$--, and $t$--independent
\be
\Omega^2=\frac{4\pi G}{d} r(x),
\ee
and interprets $G$ as the Newtonian constant and $r(x)$ as the mass density. Then the Einstein equations (\ref{EE1}) reproduce the Newtonian equation for the gravitational potential
\be
\Delta U(x)=4 \pi G r(x).
\ee

\vspace{0.5cm}

\noindent
{\bf 4. Conformal mode as a scale factor in the Eisenhart metric}\\

\noindent
Having reviewed the Eisenhart geometric description of conservative mechanical systems, let us discuss its possible generalization to encounter systems like (\ref{Osc}).
Given the form of the $x$--orbit and the impact of $\rho(t)$ upon it, a natural question arises whether a geometric description of (\ref{Osc}) exists in which the conformal mode plays the role of a cosmic scale factor.

Consider the following generalization of the Eisenhart metric (\ref{metric}):
\be\label{metr}
g_{\mu\nu}(y) dy^\mu dy^\nu=-\frac{\gamma^2 x_i x_i}{{\rho(t)}^2} dt^2- dt ds+{\rho(t)}^2 dx_i dx_i+2 q x_i d t dx_i,
\ee
where $\gamma$, $q$ are constants. A few comments are in order.

Firstly, for a fixed value of $t$ the line element in the $d$--dimensional slice parametrized by $x_i$ is given by ${\rho(t)}^2 dx_i dx_i$. Hence ${\rho(t)}^2$ may be viewed as a cosmic scale factor.

Secondly, the metric admits a covariantly constant null Killing vector field realized as in (\ref{Xi}) and hence belongs to the Kundt class. Choosing $\Omega(y)$ in Eq. (\ref{EMT}) in the form
\be
\Omega(y)=\frac{1}{L},
\ee
where $L$ is a positive constant, and imposing the Einstein equations (\ref{EE1})
one obtains the restriction on the scale factor
\be\label{RSF}
\frac{d^2 \rho}{dt^2}=\frac{\gamma^2}{\rho^3}-\frac{\rho}{L^2},
\ee
which precisely coincides with the rightmost equation in (\ref{Osc}). To the best of our knowledge, this is the first example in the literature that the conformal mechanics of de Alfaro, Fubini, and Furlan \cite{DFF} shows up in the general relativistic context.

Thirdly, computing the Christoffel symbols
\begin{align}
&
\Gamma^{s}_{tt}=-\frac{2 \gamma^2 x_i x_i(\rho \dot\rho-q)}{\rho^4}, && \Gamma^{s}_{t i}=\frac{2 x_i(q \rho \dot\rho+\gamma^2)}{\rho^2}, && \Gamma^{s}_{ij}=2(\rho\dot\rho-q)\delta_{ij},
\nonumber\\[2pt]
&
\Gamma^{i}_{tt}=\frac{\gamma^2 x_i}{\rho^4}, && \Gamma^{i}_{t j}=\frac{\dot\rho}{\rho}\delta_{ij}, &&
\end{align}
where the dot designates the derivative with respect to $t$, and analyzing the geodesic equations, one concludes that $t$ is affinely related to the proper time $\tau$
\be
\frac{dt}{d\tau}=\kappa,
\ee
where $\kappa$ is a positive constant,
the evolution of $x_i(t)$ is determined by the equation which precisely coincides with the leftmost equation in (\ref{Osc}), while $s$ is fixed from the condition that the geodesic is
null or time--like
\be\label{eqs}
\dot s=-\frac{\gamma^2 x_i x_i}{{\rho}^2}+{\rho}^2 {\dot x}_i {\dot x}_i+2 q x_i {\dot x}_i -\frac{\epsilon}{\kappa^2},
\ee
where $\epsilon=0$ for the null geodesics and $\epsilon=-1$ for the time--like geodesics.
Given the general solution in (\ref{sol}), the dynamics of the extra variable $s$ is unambiguously fixed by the first order ordinary differential equation (\ref{eqs}). Thus the original dynamical system (\ref{Osc}) is recovered if one implements the null reduction along $s$.

Note that, as compared to the conventional Eisenhart prescription, in which coordinates parametrizing the spacetime are associated with degrees of freedom of the original dynamical system, the conformal mode $\rho(t)$ enters the metric (\ref{metr}) as a specific scale factor whose time evolution is governed by the Einstein equations.

Concluding this section, we note that the $L\to\infty$ limit of (\ref{Osc}) yields a dynamical system enjoying the Schr\"odinger symmetry \cite{GM}. Curiously enough, its geometric description based upon (\ref{metr}) fails as the corresponding Riemann tensor turns out to be vanishing.

\vspace{0.5cm}

\noindent
{\bf 5. Isometries of the metric }\\

\noindent
Having fixed the form of the metric, let us establish its isometry group. The conventional way is to consider the infinitesimal transformations
\be\label{TR}
t'=t+\alpha(t,s,x), \qquad s'=s+\beta(t,s,x), \qquad x'_i=x_i+\gamma_i (t,s,x)
\ee
and demand the form of (\ref{metr}) to be intact. This yields a coupled set of first order partial differential equations for $\alpha(t,s,x)$, $\beta(t,s,x)$, and $\gamma(t,s,x)$ whose general solution determines the Killing vector fields
\bea\label{KV}
&&
H=\partial_t-x_i x_i \left(\rho^2 {\left(\frac{\dot\rho}{\rho}\right)}^{.}+2q \frac{\dot\rho}{\rho} \right) \partial_s-\frac{\dot\rho}{\rho} x_i \partial_i, \qquad \quad \qquad S=\partial_s,
\nonumber\\[2pt]
&&
P_i=\frac{1}{\rho} \cos{\left(t/L\right)} \partial_i+2 x_i \left(\frac{1}{\rho} \cos{\left(t/L\right)} (q-\rho\dot\rho)-\frac{\rho}{L} \sin{\left(t/L\right)} \right) \partial_s,
\nonumber\\[2pt]
&&
K_i=\frac{1}{\rho} L \sin{\left(t/L\right)} \partial_i+2 L x_i \left(\frac{1}{\rho} \sin{\left(t/L\right)} (q-\rho\dot\rho)+\frac{\rho}{L} \cos{\left(t/L\right)} \right) \partial_s.
\eea
The metric is also invariant under $SO(d)$ rotations acting upon $x_i$. It is straightforward to verify that $H$ is time--like, $P_i$, and $K_i$ are space--like, while $S$ is null and covariantly constant.
Computing the commutators of the vector fields, one finds
\be\label{NHA}
[H,K_i]=P_i, \qquad [H,P_i]=-\frac{1}{L^2} K_i,  \qquad [P_i,K_j]=2 S \delta_{ij}.
\ee
This is a variant of the Newton--Hooke algebra \cite{BL}~\footnote{For a more detailed discussion of the Newton--Hooke symmetry and its conformal extension see \cite{GP,AG}.} in which the covariantly constant null vector field $S$ plays the role of the central element. As $\rho(t)$ in (\ref{metr}) is treated as a fixed function, it does not come as a surprise that the $SO(2,1)$--invariance of the original dynamical system (\ref{Osc}) is not inherited by the metric (\ref{metr}).

Interestingly enough, the Newton--Hooke symmetry (\ref{KV}) provides another way of obtaining the restriction (\ref{RSF}) upon the scale factor $\rho(t)$. Indeed, the vector fields (\ref{KV}) prove to obey the structure relations (\ref{NHA}) without imposing any constraint on $\rho(t)$. Considering the metric (\ref{metr}) with arbitrary $\rho(t)$ and requiring it to admit the Killing vector fields (\ref{KV}), one immediately gets (\ref{RSF}) from the Killing equation.

The fact that (\ref{metr}) admits the time--like Killing vector field $H$, provided $\rho(t)$ obeys (\ref{RSF}), implies the spacetime is stationary. To put it in other words, there exists a coordinate system in which the metric does not depend on time. This suggests yet another possibility to arrive at Eq. (\ref{RSF}) within the geometric framework. Consider the metric (\ref{metr}) with arbitrary scale factor $\rho(t)$ and implement the coordinate transformation
\be\label{TRF}
t'=t, \qquad x'_i=\rho(t) x_i, \qquad s'=s+x_i x_i (\rho\dot\rho-q),
\ee
which brings it to the form
\be\label{MOsc}
g'_{\mu\nu}(y') dy'^\mu dy'^\nu=-\left(\frac{\gamma^2}{\rho^4}-\frac{\ddot\rho}{\rho} \right) x'_i x'_i dt'^2- dt' ds'+dx'_i dx'_i.
\ee
Requiring the resulting metric to be Lorentzian and stationary, one gets
\be\label{MCON}
\frac{\gamma^2}{\rho^4}-\frac{\ddot\rho}{\rho}=\frac{1}{L^2},
\ee
where $L$ is constant,
which reproduces (\ref{RSF}). Note that, if (\ref{MCON}) is satisfied, (\ref{MOsc}) gives the Eisenhart metric associated with the isotropic oscillator. The coordinate system (\ref{TRF}) is thus
the analog of the non--inertial Newton--Hooke reference frame (\ref{TRA}).

\vspace{0.5cm}

\noindent
{\bf 6. Conclusion }\\

\noindent
To summarize, in this work we proposed a geometric formulation for a particular dynamical system which describes the isotropic oscillator in arbitrary dimension driven by the one--dimensional conformal mode.
In contrast to the conventional Eisenhart prescription, in which coordinates parametrizing the spacetime are associated with degrees of freedom of the original dynamical system, the conformal mode enters the metric as a specific scale factor. The equation which governs its evolution was obtained in three different ways either by imposing the Einstein equations, or demanding the Newton--Hooke isometry group, or requiring the spacetime to be stationary. To the best of our knowledge, the consideration above provides the first example in the literature that the conformal mechanics of de Alfaro, Fubini, and Furlan shows up in the general relativistic context.

Turning to possible further development, it would be interesting to generalize this work to the specific chain of isotropic oscillators driven by $d=1$ conformal mode which enjoys
the so called $l$--conformal Newton--Hooke symmetry \cite{GM}.

\vspace{0.5cm}

\noindent{\bf Acknowledgements}\\

\noindent
This work was supported by the Tomsk Polytechnic University competitiveness enhancement program. We thank S. Krivonos for communicating to us that the system (\ref{Osc}) is Lagrangian.

\end{document}